\documentclass[aps,twocolumn,superscriptaddress,showpacs]{revtex4}%

\usepackage{amssymb}
\usepackage{amsmath}
\usepackage{graphicx}
\usepackage[normalem]{ulem}
\usepackage[dvips]{color}

\usepackage[normalem]{ulem} 
\usepackage[dvips]{color} 
\renewcommand\sout{\bgroup \color{red} \ULdepth=-.5ex \ULset}

\def\esym{$E_{sym}(\rho)$~}
\def\rpi {$\pi^-/\pi^+$~}
\def\es0{$E_{sym}(\rho_0)$~}
\begin{document}

\title{Elimination of influence of neutron-skin size difference of initial colliding nuclei in Pb+Pb collisions}
\author{Gao-Feng Wei}\email[Email address: ]{wei.gaofeng@foxmail.com}
\affiliation{Department of Applied Physics, Xi'an JiaoTong University, Xi'an 710049, China}
\affiliation{School of Physics and Mechatronics Engineering, Xi'an University of Arts and Science, Xi'an, 710065, China}
\affiliation{Department of Physics and Astronomy, Texas A$\&$M University-Commerce, Commerce, TX 75429-3011, USA}

\begin{abstract}
Within an isospin- and momentum-dependent transport model
using as an input nucleon density profiles from Hartree-Fock
calculations based on a modified Skyrme-like (MSL) model, we study
how to eliminate the influence of neutron-skin size difference
of initial colliding nuclei in probing the nuclear symmetry energy.
Within the current experimental uncertainty range of neutron-skin
size of $^{208}$Pb, the Pb+Pb collisions are performed in semicentral
and peripheral collisions with impact parameters of 5 and 9fm and
at beam energies from 50 MeV/nucleon to 1000 MeV/nucleon, respectively.
It is shown that combination of neutron and proton collective flows,
i.e., neutron-proton differential elliptic flow, neutron-proton elliptic
flow difference, neutron-proton differential transverse flow and
neutron-proton transverse flow difference, can effectively eliminate
the effects of neutron-skin size difference and thus can be as useful
sensitive observables in probing nuclear matter symmetry energy
in heavy-ion collisions. Moreover, the combined neutron-proton stopping
power including the neutron-proton differential stopping power and
neutron-proton stopping power difference can also eliminate the effects
of neutron-skin size difference and shows some sensitivities to symmetry
energy especially at the lower beam energy.
\end{abstract}


\maketitle

\section{Introduction}\label{introduction}
The density dependence of nuclear symmetry energy \esym as one of
important issue in isospin physics has been studied for the past few
decades because of its very importance not only in understanding
the structure of radiative nuclei in nuclear physics
\cite{Oya98,Bro00,Hor01,Fur02} but also its crucial roles in nuclear
astrophysics \cite{Bet90,Lat01,Eng94,Pra88,Che03a}.
Up to now, although many useful experimental observables \cite{Bao04,Mul95,
Bar02,Che03a,Che03b,Shi03,Sca99,Bao01,Bao02,Gai04,Dan98} have been proposed
to determine the nuclear symmetry energy in heavy-ion collisions,
they have had limited success because of its sensitivity not only to
nuclear symmetry energy but also other poorly known physical effects
\cite{Ditoro05,Bao08}. Nevertheless, a key step in determining the nuclear symmetry
energy is the determination of experimental observables which can be
serve as clean and sensitive probes \cite{Ditoro05,Bao08}. Therefore, the search
of the experimental observables sensitive only to nuclear symmetry
energy than others is a crucial task in determination of nuclear symmetry energy.

It is well known that heavy-ion reactions induced by neutron-rich nuclei
provide an important opportunity to constrain the symmetry energy in a broad
density range \cite {Sto86,Cas90,Har96,Ko96,Bao98,Ditoro99}. To initialize transport
models of heavy-ion reactions, it is necessary to know the nucleon density profiles
for the two colliding nuclei. Generally speaking, one should use the Thomas-Fermi method to
extract the density profiles for the colliding nuclei. Practically, one usually
use other methods to approximate Thomas-Fermi method during initializing the
colliding nuclei such as the droplet model used in the isospin-dependent quantum
molecular dynamics (IQMD) model \cite{Sun10,Dai14} and the Skyrme Hartree-Fock model
used in the isospin-dependent Boltzmann-Uehling-Uhlenbeck (IBUU) model \cite{IBUU}.
In any case, the initialization should guarantee the numerical stabilization
of colliding nuclei and fit the bulk nuclei properties as good as possible; see, e.g.,
Refs.\cite {Dan84,Bot90}. The main problem of these methods is the numerical
stability of initial colliding nuclei in the subsequent reaction. However, this allows one
to easily separate effects on the final observables due to the initial state from those
due to the reactions \cite{Wei13}. Based on this consideration, by examining the relative
effects of neutron-skin size in initial nuclei and the symmetry energy at suprasaturation
densities reached in heavy-ion collisions on the charged pion ratio in the final state,
we have shown recently that the \rpi ratio is sensitive not only to nuclear symmetry energy
but also the neutron-skin size of initial colliding nuclei especially in peripheral collisions,
see Ref.~\cite{Wei13}. Nevertheless, as mentioned above a crucial task in determination of
symmetry energy is to find the clean experimental observables. Under this consideration in mind,
we are going to find which experimental observables are sensitive to nuclear symmetry energy
than neutron-skin size difference of initial colliding nuclei. To this end, in this work we
investigate how to eliminate the influence of the neutron-skin size difference of initial
colliding nuclei in probing the nuclear symmetry energy in Pb+Pb heavy-ion collisions.
It can be found later that the combined neutron-proton collective flow and stopping power
can effectively eliminate the effects of neutron-skin size difference of initial colliding nuclei
but keep the effects of symmetry energy especially at the lower beam energy.

\section{the model}
In this part we briefly describe the model used in the present
study, i.e., the isospin- and momentum-dependent
Boltzmann-Uehling-Uhlenbeck transport model \cite{IBUU} of
version IBUU11 \cite{Ouli}. The momentum dependence of both the
isoscalar \cite{GBD87,PDG88,MDYI90,Dan00,Greco99} and isovector
\cite{IBUU,Das03,Chen04,Rizzo04} parts of the nuclear interaction
is important in understanding not only many phenomena in
intermediate-energy heavy-ion collisions but also thermodynamical
properties of isospin-asymmetric nuclear matter \cite{Xuj07a,Xuj07b,Xuj08}.
The mean-field potential for a nucleon with momentum $\vec{p}$ and
isospin $\tau$ can be written as \cite{Das03}
\begin{eqnarray}
U(\rho,\delta ,\vec{p},\tau ) &=&A_{u}(x)\frac{\rho _{-\tau }}{\rho _{0}}%
+A_{l}(x)\frac{\rho _{\tau }}{\rho _{0}}  \notag \\
&+&B(\frac{\rho }{\rho _{0}})^{\sigma }(1-x\delta ^{2})-8\tau x\frac{B}{%
\sigma +1}\frac{\rho ^{\sigma -1}}{\rho _{0}^{\sigma }}\delta \rho
_{-\tau }
\notag \\
&+&\frac{2C_{\tau ,\tau }}{\rho _{0}}\int d^{3}p^{\prime }\frac{f_{\tau }(%
\vec{p}^{\prime })}{1+(\vec{p}-\vec{p}^{\prime })^{2}/\Lambda ^{2}}
\notag \\
&+&\frac{2C_{\tau ,-\tau }}{\rho _{0}}\int d^{3}p^{\prime }\frac{f_{-\tau }(%
\vec{p}^{\prime })}{1+(\vec{p}-\vec{p}^{\prime })^{2}/\Lambda ^{2}}.
\label{MDIU}
\end{eqnarray}%
In the above, $\rho=\rho_n+\rho_p$ is the nucleon number density and
$\delta=(\rho_n-\rho_p)/\rho$ is the isospin asymmetry of the
nuclear medium; $\rho_{n(p)}$ denotes the neutron (proton) density,
isospin $\tau$ is $1/2$ for neutrons and $-1/2$ for protons, and
$f(\vec{p})$ is the local phase space distribution function. The
expressions and values of the parameters $A_{u}(x)$, $A_{l}(x)$,
$\sigma$, $B$, $C_{\tau ,\tau }$, $C_{\tau ,-\tau }$, and $\Lambda $
can be found in Refs. \cite{Das03,Che05}, and they lead to the
binding energy of $-16$ MeV, incompressibility $212$ MeV for
symmetric nuclear matter, and symmetry energy $E_{sym}(\rho_0)=30.5$
MeV at saturation density $\rho_0=0.16$ fm$^{-3}$, respectively.

\begin{figure}[h]
\centerline{\includegraphics[scale=0.30]{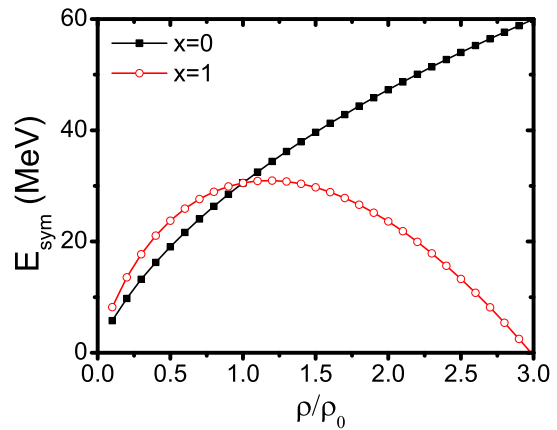}} \caption{(Color
online) The density dependence of the symmetry energy. Taken from
Ref. \cite{Wei13}.} \label{Esym}
\end{figure}

The variable $x$ is introduced to mimic different forms of the
symmetry energy predicted by various many-body theories without changing
any properties of symmetric nuclear matter and the value of $E_{sym}(\rho_0)$.
At suprasaturation densities, although the IBUU calculations favour by comparing
with the FOPI data a super soft symmetry energy with $x=1$ \cite{Xiao09},
compared to the FOPI data the ImIQMD calculations by Feng $et~al$ \cite{Feng10}
show a super hard symmetry energy. Therefore, to evaluate the relative effects
of symmetry energy we use two values of $x=1$ and $x=0$ as the so-called soft
and stiff symmetry energy parameters as shown in Fig. \ref{Esym} \cite{Wei13}.
It should be mentioned that the current uncertain range of symmetry energy at
suprasaturation densities is much larger than the one used here \cite{Bao08,Xuj12}.
The density dependence of \esym around $\rho_0$ is generally characterized by
the slope parameter $L\equiv 3\rho_{0}(dE_{sym}/d\rho) _{\rho=\rho_{0}}$.
The softer (stiffer) \esym with $x=1$ ($x=0$) has a value of $L=16.4$ (62.1)
MeV.
\begin{figure}[h]
\centerline{\includegraphics[scale=0.30]{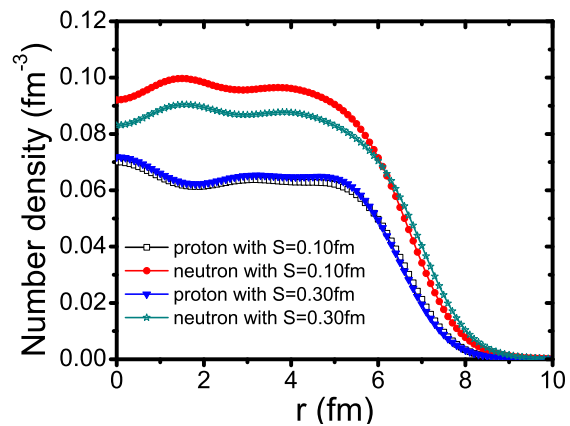}} \caption{(Color
online) The neutron and proton density profiles for $^{208}$Pb with
neutron-skin thickness of 0.1 and 0.3 fm, respectively.
Taken from Ref. \cite{Wei13}.} \label{Density}
\end{figure}
\begin{figure}[h]
\centerline{\includegraphics[scale=0.30]{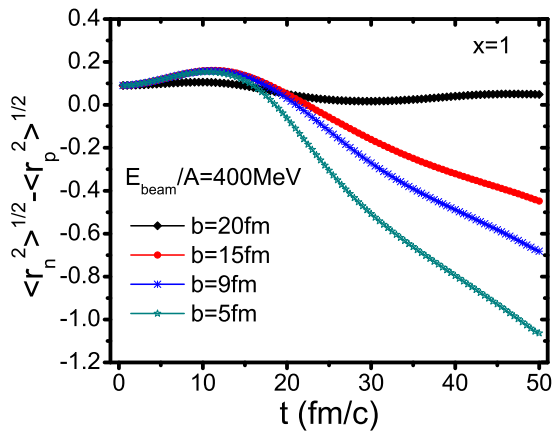}} \caption{(Color
online) Evolution of the rms radius difference between neutron and
proton initially setting as 0.1fm in $^{208}$Pb+$^{208}$Pb collisions
with impact parameters of 5, 9, 15 and 20fm at beam energy of
400MeV/nucleon, respectively.}\label{gscheck}
\end{figure}

To examine effects of the neutron-skin size of initial colliding nuclei,
we initialize nucleons in phase space using neutron and proton density
profiles predicted by Hartree-Fock calculations based on the MSL
model \cite{Che09,Che10}. Different values of neutron-skin thickness can be
obtained by changing only the value of $L$ in the MSL0 force~\cite{Che10} while
keeping all the other macroscopic quantities the same. Shown in Fig.
\ref{Density} are the density profiles corresponding to a neutron-skin
thickness $S$ of 0.1 and 0.3 fm of $^{208}$Pb \cite{Wei13}, which are
in the range of about $0.11\pm 0.06$ fm from $\pi^+$-Pb scattering~\cite{frie} to
$0.33^{+0.16}_{-0.18}$ fm from the PREX-1 experiments using parity
violating e-Pb scattering \cite{Prex1}. Although these available data suffer
from large uncertainties, it was shown very recently within a relativistic
mean-field model \cite{Farrooh} that a neutron-skin for $^{208}$Pb
as thick as $0.33+0.16$ fm reported by the PREX-I experiment \cite{Prex1}
can not be ruled out although most other studies have reported much
smaller average values albeit largely overlapping with the PREX-I result
within error bars. It is expected that the proton distributions are almost
identical, while the neutrons distribute differently in the two cases
considered.

Before showing the results of the studies, one must check the stability
of the ground state nucleus. To this end, one can check the time evolution
of the rms radius difference between neutrons and protons in the projectile
and/or target with impact parameter to be infinite in the reaction model.
This is because when the impact parameter is infinite, the projectile and
target can not touch with each other and the corresponding interaction
between them becomes zero, they are in their ground states and move
along their initial trajectory; it is naturally that the initial rms
radius difference between neutrons and protons remains unchanged. Shown in
Fig. \ref{gscheck} is the evolution of the rms radius difference between
neutrons and protons initially setting as 0.1fm both in the projectile and
target for different impact parameters at beam energies of 400MeV/nucleon,
respectively. Firstly, it can be found that the value of the rms radius
difference between neutrons and protons decreases rapidly with the impact
parameter decreasing due to the interactions between the projectile and target
as well as the corresponding collisions among nucleons increasing. Secondly,
it is expected that the initial rms radius difference between neutrons and
protons in the projectile and/or target approximately approaches stable as the
impact parameter increasing to be 20fm. Certainly, a long-period and
small-amplitude oscillation of the rms radius difference is still seen due to
the distance between the projectile and target inadequate far and the tiny
fluctuation from collisions between nucleons in the same nucleus.
However, this level of stability of the ground state nucleus more or less
lasts long enough, which should be reflected in the final reaction production.

\section{Results and discussions}\label{results}
We now present results of the study in the following. Considering that
there is no obvious neutron-skin effects in head-on heavy-ion reactions
as shown in our previous work \cite{Wei13}, we thus carry out this study
in semicentral and peripheral $^{208}$Pb+$^{208}$Pb collisions with
impact parameters of 5 and 9fm. On the other hand, to distinguish the
difference of neutron-skin size of $S$=0.1 and 0.3fm, we have performed
large scale calculations with $4\times 10^5$ events in each case reported
here. Thus, the statistical error bars are smaller than the plotting
symbols in most plots. In addition, the percentage used in some
figure is to avoid too much numerical digit after the decimal.

\subsection{The elliptic flow}

\begin{figure}[h]
\centerline{\includegraphics[scale=0.41]{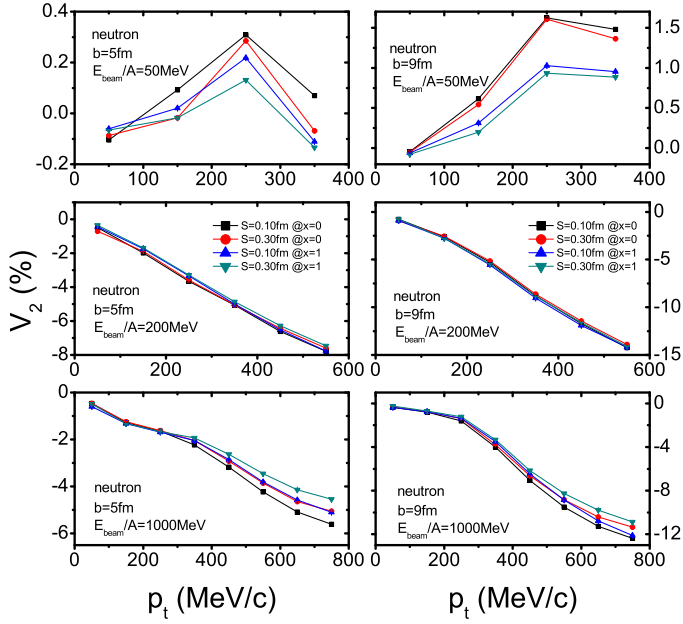}} \caption{(Color
online) The transverse momentum dependence of elliptic flow of the
midrapidity ($|y/y_{beam}|\leq 0.5$) neutrons in semicentral and peripheral
$^{208}$Pb+$^{208}$Pb collisions with impact parameter of 5 and 9fm and at
beam energies from 50 to 1000 MeV/nucleon, respectively.} \label{ptv2n}
\end{figure}

\begin{figure}[h]
\centerline{\includegraphics[scale=0.41]{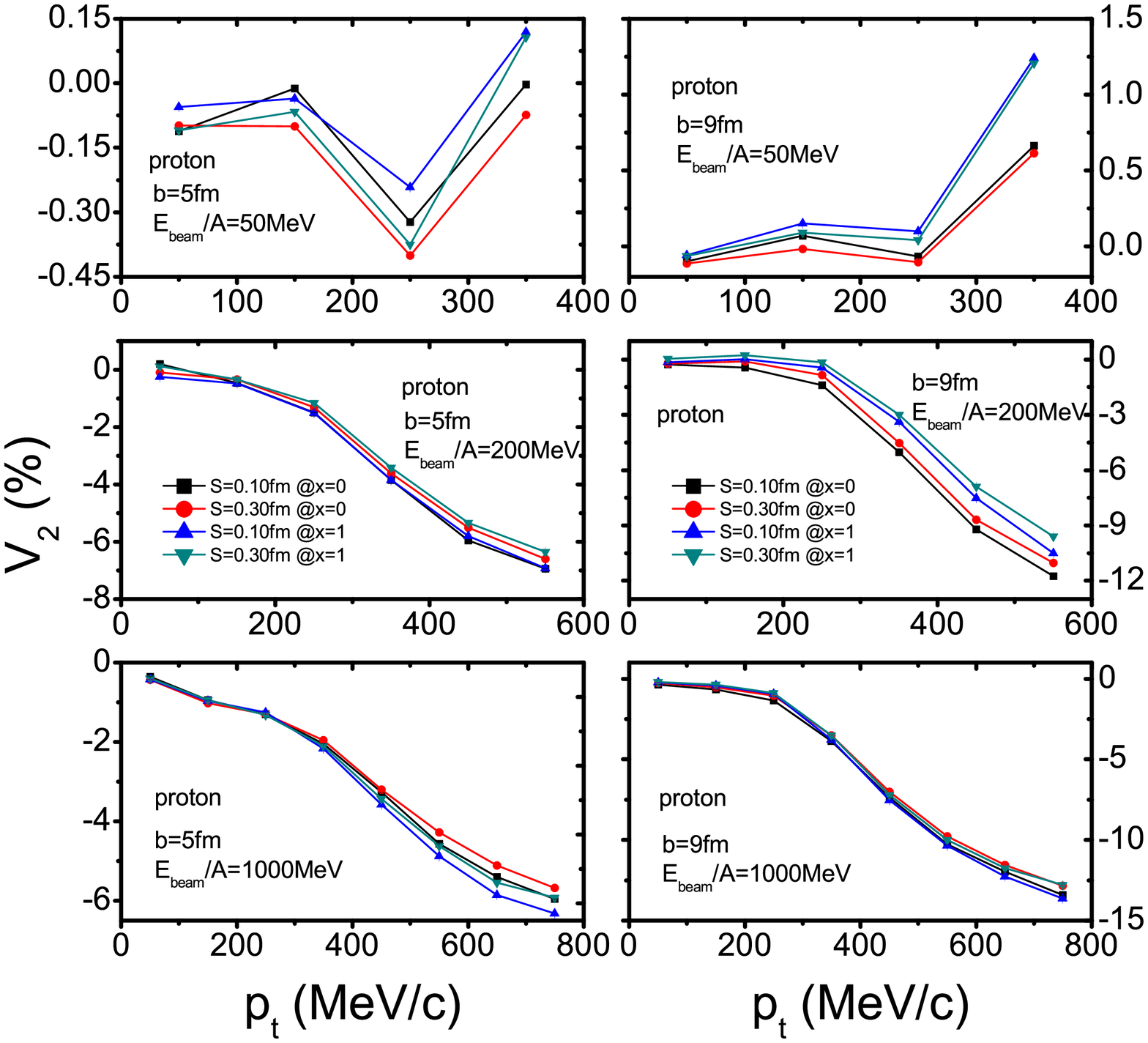}} \caption{(Color
online) Same as Fig. \ref{ptv2n} but for protons.} \label{ptv2p}
\end{figure}

The elliptic flow has been widely used to study the properties of the hot and
dense matter formed in the early stage of heavy-ion collisions at relativistic
and intermediate energies, see, e.g., Refs. \cite{Song11,Dan02a,Xuj12}. To probe
the density dependence of symmetry energy, we show in Fig. \ref{ptv2n} and
Fig. \ref{ptv2p} transverse momentum dependence of elliptic flow of the
midrapidity ($|y/y_{beam}|\leq 0.5$) neutrons and protons in semicentral and
peripheral $^{208}$Pb+$^{208}$Pb collisions with impact parameters of 5 and 9fm
and at beam energies from 50 to 1000 MeV/nucleon, respectively. It can be found
that the elliptic flow of nucleons shows a transition from in-plane to out-of-plane
with the beam energy increasing from 50 to 1000MeV/nucleon. This is because at the
lower beam energies, the mean field dominates the reaction dynamics and causes
the in-plane enhancement of emitted reaction products, while with the beam energies
increasing, the mean field becomes less important and the collective expansion
process based on nucleon-nucleon scattering starts to be predominant, and the
squeezed out elliptic flow as the result of shadowing of spectator starts to become
negative \cite{Zym99,Sina11,Trau12}. However, the symmetry energy effects on
elliptic flow of nucleons are not obvious as reported in previous
Refs. \cite{Ditoro05,Ditoro10}, and the neutron-skin effects on elliptic flow
of nucleons are comparable with or larger than that of symmetry energy.
\begin{figure}[h]
\centerline{\includegraphics[scale=0.41]{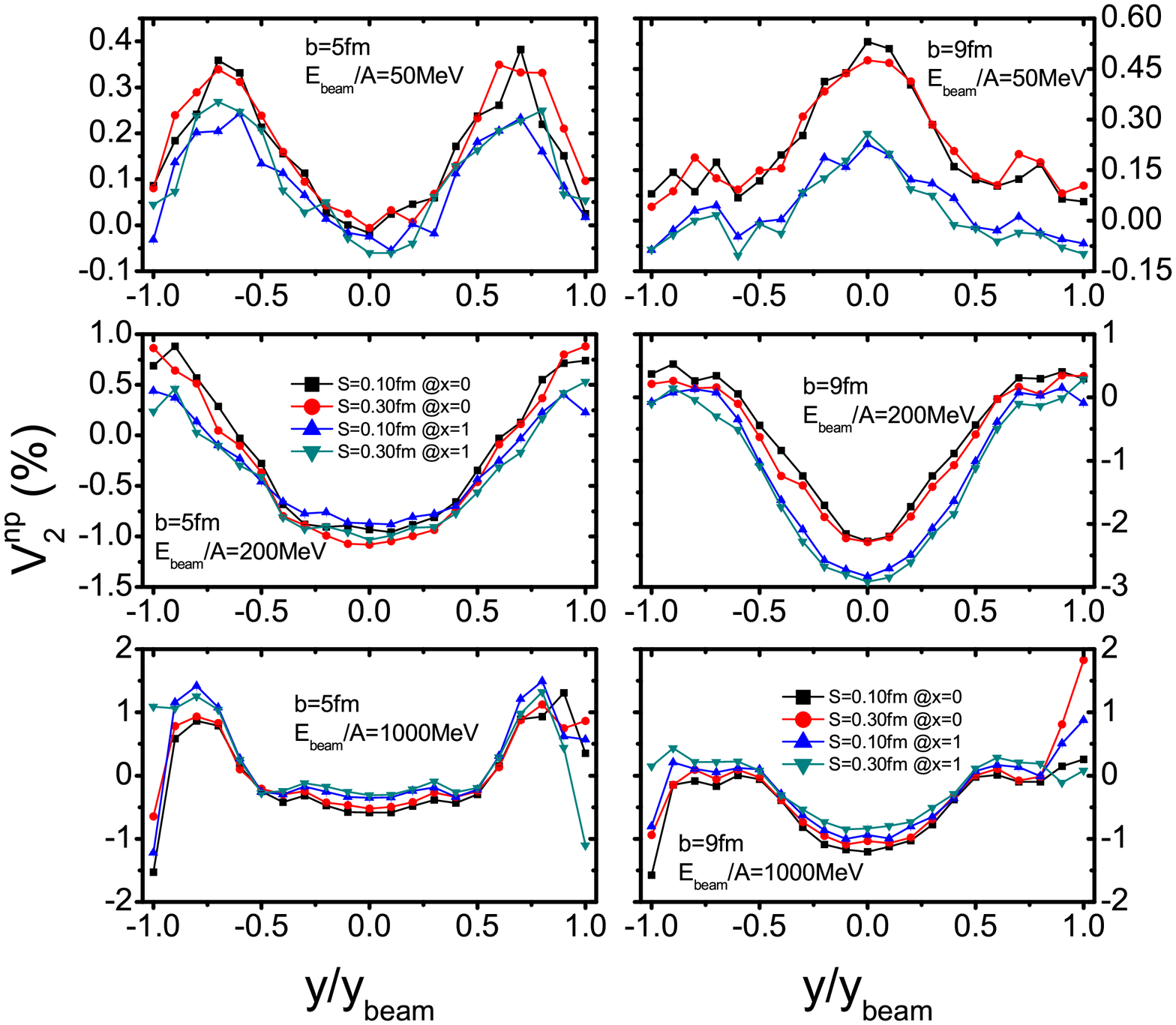}} \caption{(Color
online) The rapidity dependence of neutron-proton differential elliptic
flow in $^{208}$Pb+$^{208}$Pb collisions with impact parameters of 5 and
9fm and at beam energies from 50 to 1000 MeV/nucleon, respectively.} \label{yv2dnp}
\end{figure}
Nevertheless, the elliptic flow of both neutron and proton has larger value with thinner
neutron-skin due to more nucleons in spectator generating the stronger shadowing effects on
in-plane emitted nucleons and leading squeezed out elliptic flow of out-of-plane
nucleons to be larger. In other words, the effects of neutron-skin size on elliptic flow of
neutron and proton are approximately identical except for the additional Coulomb repulsion
between protons. This naturally leads us to check whether combining the neutron
and proton elliptic flows can eliminate the effects of neutron-skin size difference
but keep the effects of symmetry energy. To this end, we formulate the so-called
neutron-proton differential elliptic flow
\begin{eqnarray}\label{diffflow}
v_{2}^{np}(u)&=&\frac{1}{N(u)}\sum\limits_{i=1}^{N(u)}v_{2}^{i}(u)w_{i} \notag\\
&=&\frac{N_{n}(u)}{N(u)}<v_{2}^{n}(u)>-\frac{N_{p}(u)}{N(u)}<v_{2}^{p}(u)>,
\end{eqnarray}
where $N(u)$, $N_{n}(u)$ and $N_{p}(u)$ are the numbers of nucleons,
neutrons and protons at parameter $u$ which denotes the rapidity $y$
or transverse momentum $p_{t}$, and $w_{i}$ is $1$ for
neutrons and $-1$ for protons, respectively.
\begin{figure}[h]
\centerline{\includegraphics[scale=0.41]{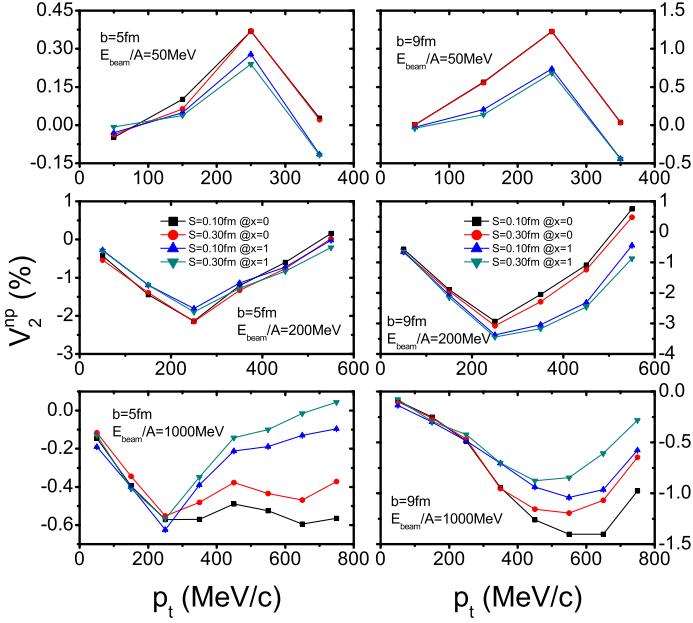}} \caption{(Color
online) The transverse momentum dependence of neutron-proton differential
elliptic flow of the midrapidity ($|y/y_{beam}|\leq 0.5$)
in $^{208}$Pb+$^{208}$Pb collisions with impact parameters of 5 and 9fm and
beam energies from 50 to 1000 MeV/nucleon, respectively.} \label{ptv2dnp}
\end{figure}
Shown in Fig. \ref{yv2dnp} is the rapidity dependence of neutron-proton differential
elliptic flow in semicentral and peripheral $^{208}$Pb+$^{208}$Pb collisions with
impact parameters of 5 and 9fm and at beam energies from 50 to 1000 MeV/nucleon,
respectively. It is expected that the neutron-proton differential elliptic flow
shows obvious sensitivity to the symmetry energy. This can be understandable since
the combined elliptic flow constructively maximizes the effects of the symmetry
potential but minimizes the effects of the isoscalar potential similar to that of
neutron-proton differential transverse flow proposed by Li \cite{Bao00}. On the other
hand, noticing that the neutron-proton differential elliptic flow in midrapidity
($|y/y_{beam}|\leq 0.5$) is hardly affected by the neutron-skin size difference
but sensitive to symmetry energy, the transverse momentum dependence of
neutron-proton differential elliptic flow of the midrapidity ($|y/y_{beam}|\leq 0.5$)
is shown in Fig. \ref{ptv2dnp}. It can be found that the transverse momentum dependence of
neutron-proton differential elliptic flow is indeed more sensitive to symmetry energy
but hardly sensitive to the neutron-skin size difference especially at the lower beam energy.

\begin{figure}[h]
\centerline{\includegraphics[scale=0.41]{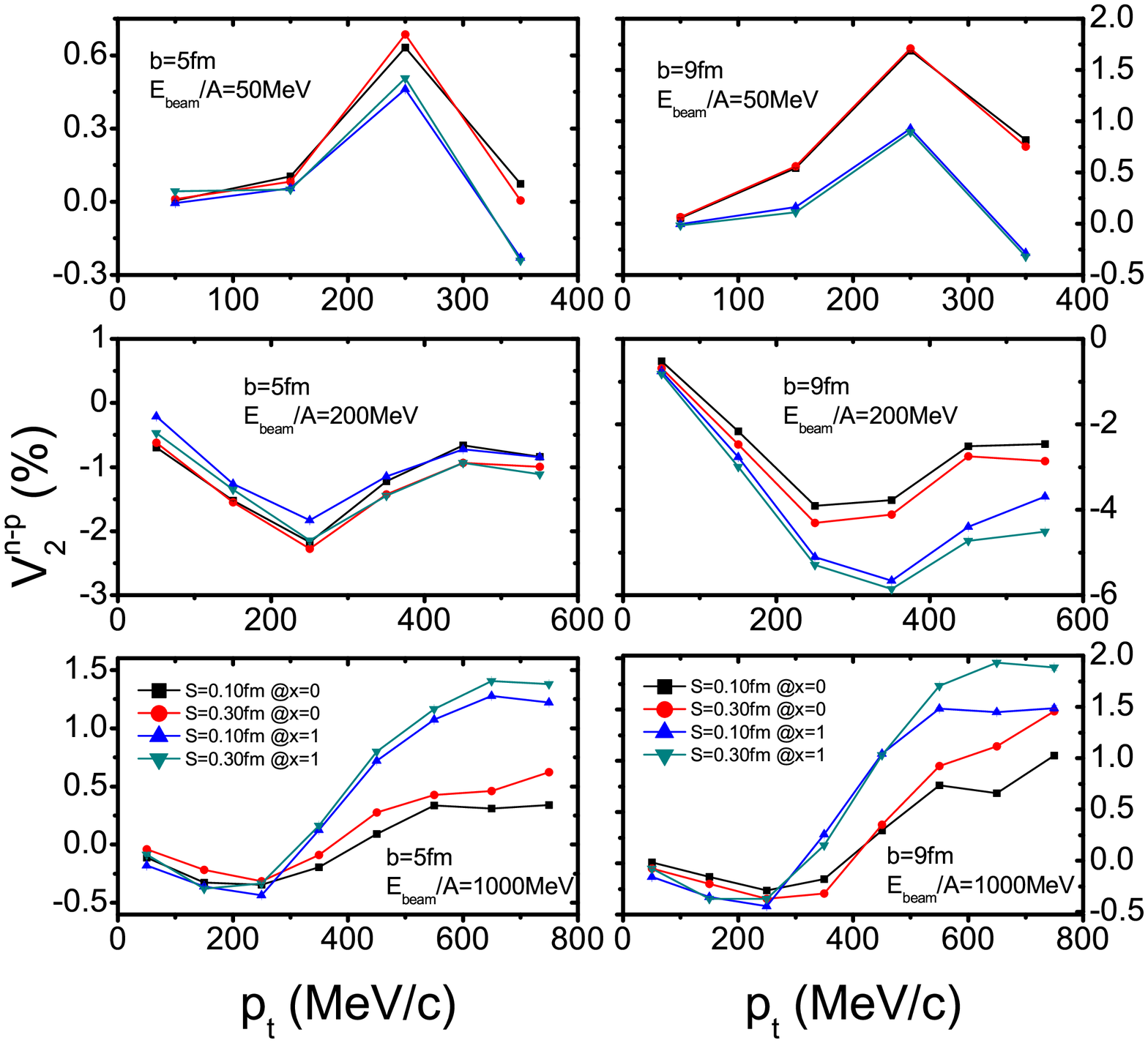}} \caption{(Color
online) The transverse momentum dependence of neutron-proton elliptic flow
difference of the midrapidity ($|y/y_{beam}|\leq 0.5$) in
$^{208}$Pb+$^{208}$Pb collisions with impact parameter of 5 and 9fm and beam energies
from 50 to 1000 MeV/nucleon, respectively.} \label{ptv2npd}
\end{figure}

Another combination of the neutron and proton elliptic flows is the direct difference
of them proposed in Ref. \cite{Ditoro05,Ditoro10} defined as
\begin{equation}
v_{2}^{n-p}(p_{t})=<v_{2}^{n}(p_{t})>-<v_{2}^{p}(p_{t})>,
\end{equation}
which should also be sensitive to symmetry energy since it is the special
case of the neutron-proton differential elliptic flow, i.e., when the neutron
and proton have the same multiplicities but different average elliptic flow.
Shown in Fig. \ref{ptv2npd} is the transverse momentum dependence of
neutron-proton elliptic flow difference of the midrapidity ($|y/y_{beam}|\leq 0.5$)
in semicentral and peripheral $^{208}$Pb+$^{208}$Pb collisions with impact parameters
of 5 and 9fm and at beam energies from 50 to 1000 MeV/nucleon, respectively.
It is seen that the transverse momentum dependence of neutron-proton elliptic flow
difference is indeed sensitive symmetry energy but hardly sensitive to the
neutron-skin size difference especially at the lower beam energy. These indicate that
combination of neutron and proton elliptic flows, i.e., neutron-proton differential
elliptic flow and neutron-proton elliptic flow difference, can effectively eliminate
the effects of neutron-skin size difference especially at the lower beam energy and thus
can be as useful sensitive observables in probing nuclear matter symmetry energy
in heavy-ion collisions.

\subsection{The transverse flow}

\begin{figure}[h]
\centerline{\includegraphics[scale=0.35]{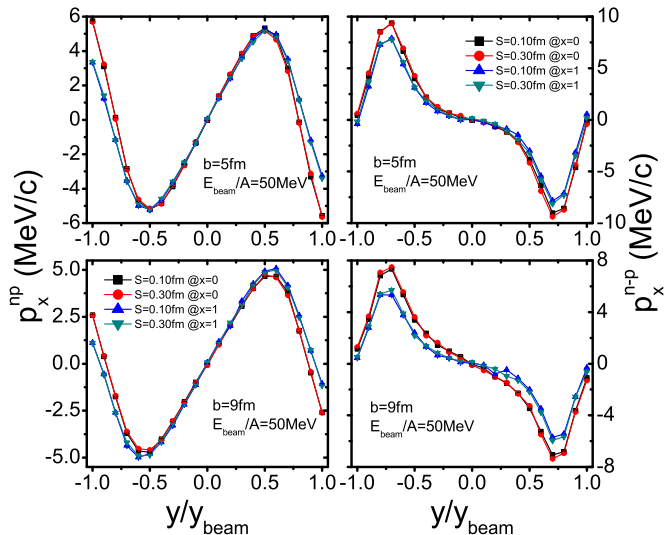}} \caption{(Color
online) The neutron-proton differential transverse flow (left panel) and neutron-proton
transverse flow difference (right panel) in $^{208}$Pb+$^{208}$Pb collisions with impact
parameter of 5 and 9fm and beam energy of 50MeV/nucleon, respectively.}
\label{pxnp}
\end{figure}

The neutron-proton differential transverse flow as a good tracer of the
symmetry potential defined as
\begin{eqnarray}\label{px}
p_{x}^{np}(y)&=&\frac{1}{N(y)}\sum\limits_{i=1}^{N(y)}p_{x}^{i}(y)w_{i} \notag\\
&=&\frac{N_{n}(y)}{N(y)}<p_{x}^{n}(y)>-\frac{N_{p}(y)}{N(y)}<p_{x}^{p}(y)>,
\end{eqnarray}
was proposed to measure symmetry energy in heavy-ion collisions
due to its advantages of combining constructively effects of the symmetry
potential on the isospin fractionation and the collective flow \cite{Bao00}.
When the neutron and proton have the same multiplicities but different average
transverse flow, the neutron-proton differential transverse flow naturally becomes
the direct difference of the neutron and proton transverse flows, i.e., the
neutron-proton transverse flow difference,
\begin{eqnarray}\label{px}
p_{x}^{n-p}(y)=<p_{x}^{n}(y)>-<p_x^{p}(y)>.
\end{eqnarray}
Similar to the neutron-proton differential elliptic flow and neutron-proton elliptic
flow difference, combination of the neutron and proton transverse flows, i.e.,
neutron-proton differential transverse flow and neutron-proton transverse flow
difference, can effectively eliminate the effects of neutron-skin size difference but
keep the effects of symmetry energy especially at the lower beam energy as shown in
Fig. \ref{pxnp}. At the beam energies of 200 and 1000MeV/nucleon, the neutron-proton
differential transverse flow and/or neutron-proton transverse flow difference
are less sensitive to nuclear symmetry energy compared to those at 50MeV/nucleon.
\begin{figure}[h]
\centerline{\includegraphics[scale=0.35]{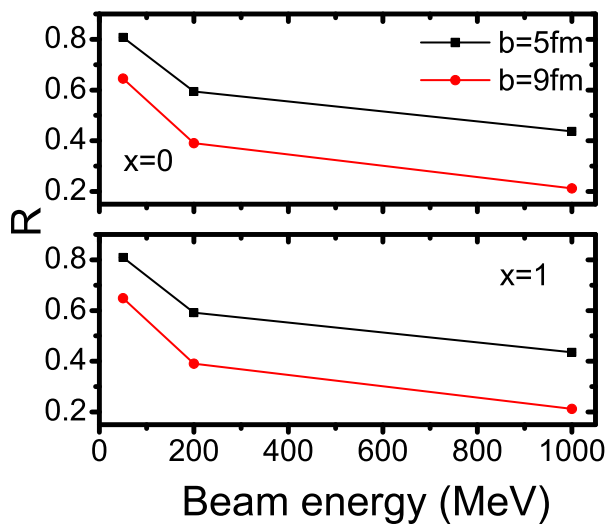}} \caption{(Color
online) The beam energy and impact parameter dependence of stopping power of all
nucleons in $^{208}$Pb+$^{208}$Pb collisions with the stiff (upper panel) and
soft (lower panel) symmetry energy, respectively.} \label{sp-global}
\end{figure}
\subsection{The stopping power}
Collective flow, generated by pressure gradient of dense nuclear matter
formed in heavy ion collision, is closely related to nuclear stopping power.
Large stopping power can lead to a remarkable pressure gradient in the
compressed nuclear matter. It is also believed that the stopping power
governs most of the dissipated energy, the amplitude of large collective motion,
the maximum attainable baryon and energy densities, as well as the thermalization
of the collision system \cite{And06,Luo07,Leh10}. Two kinds of ratio of transverse to
longitudinal quantities are commonly used to measure the degree of stopping; one is
the energy-based isotropy ratio, another is the momentum-based isotropy ratio.
They are actually the different forms of physics realization of the classical
Maxwell distribution assumption \cite{Zhang11}. Here, the momentum-based
isotropy ratio is employed to measure the degree of stopping, its definition
is
\begin{eqnarray}\label{R}
R=\frac{2}{\pi}\frac{\sum |p_{ti}|}{\sum |p_{zi}|},
\end{eqnarray}
where $p_{ti}$($p_{zi}$) is the transverse (longitudinal) momentum of nucleon in center of
mass system, and the sum runs over all products event by event. It is believed that
full stopping is reached when the ratio $R$ reaches the value of 1 \cite{Str83};
and the superstopping of the ratio $R>1$ is explained by the preponderance of momentum
flow perpendicular to the beam direction \cite{Ren84}.
\begin{figure}[h]
\centerline{\includegraphics[scale=0.41]{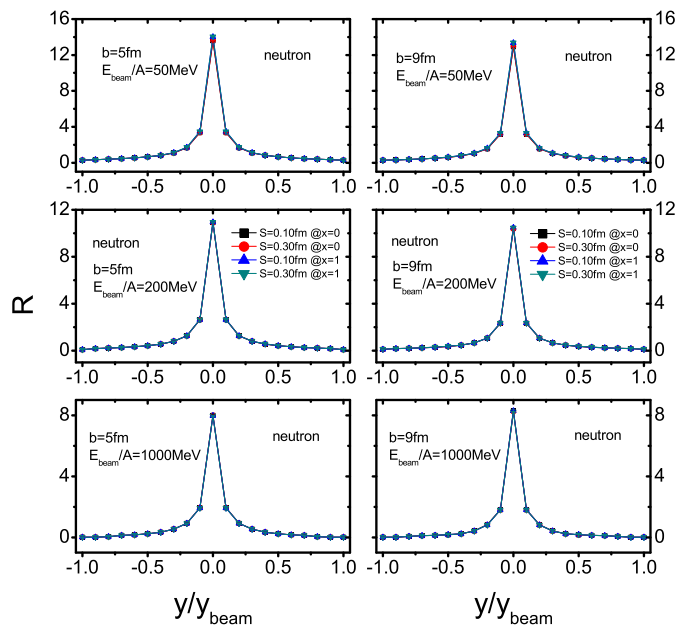}} \caption{(Color
online) The rapidity dependence of stopping power of neutrons in $^{208}$Pb+$^{208}$Pb collisions
with impact parameter of 5 and 9fm and beam energies from 50 to 1000 MeV/nucleon, respectively.}
\label{nsp-ry}
\end{figure}
\begin{figure}[h]
\centerline{\includegraphics[scale=0.41]{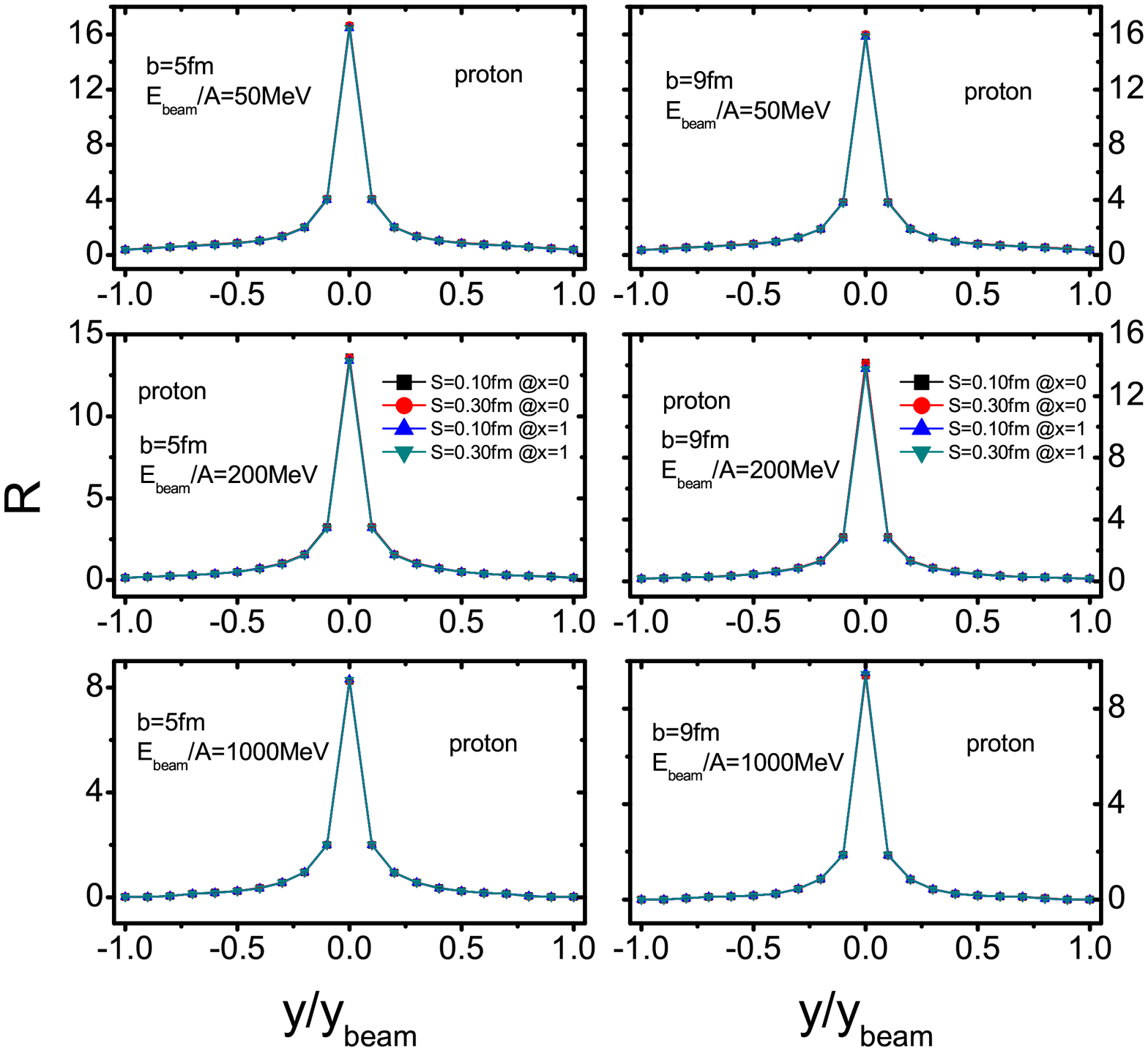}} \caption{(Color
online) Same as Fig. \ref{nsp-ry} but for protons.} \label{psp-ry}
\end{figure}
\begin{figure}[h]
\centerline{\includegraphics[scale=0.41]{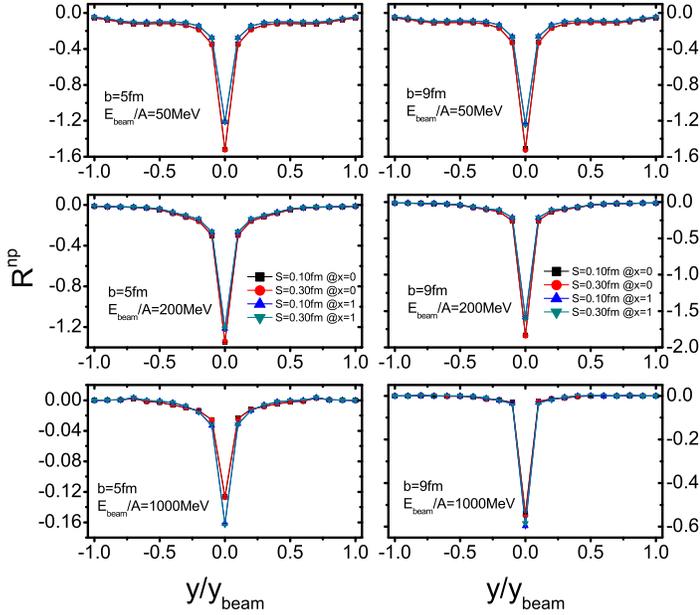}} \caption{(Color
online) The rapidity dependence of neutron-proton differential stopping power in $^{208}$Pb+$^{208}$Pb
collisions with impact parameter of 5 and 9fm and beam energies from 50 to 1000 MeV/nucleon,
respectively.} \label{npsp-ry}
\end{figure}
\begin{figure}[h]
\centerline{\includegraphics[scale=0.40]{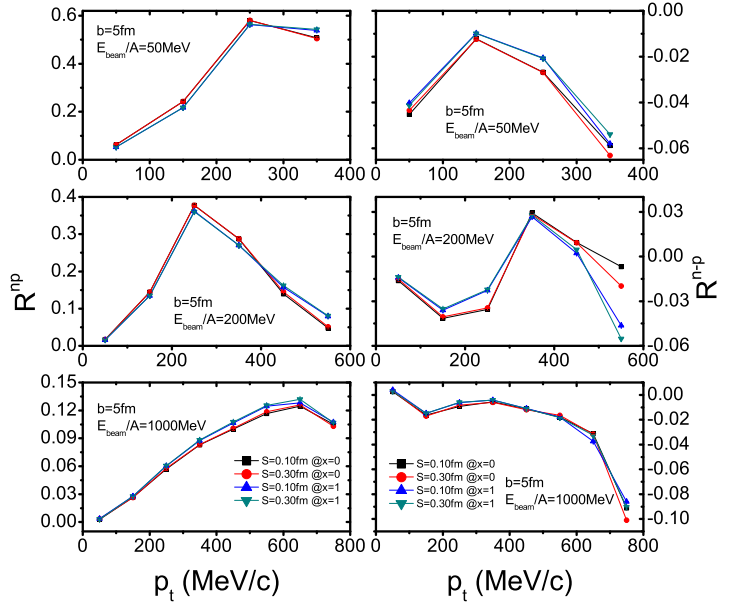}} \caption{(Color
online) The transverse momentum dependence of neutron-proton differential
stopping power (left panel) and neutron-proton stopping power difference (right panel)
of the midrapidity ($|y/y_{beam}|\leq 0.5$) in $^{208}$Pb+$^{208}$Pb collisions
with impact parameter of 5fm and beam energies from 50 to 1000MeV/nucleon, respectively.}
\label{npsp-pt1}
\end{figure}
\begin{figure}[h]
\centerline{\includegraphics[scale=0.40]{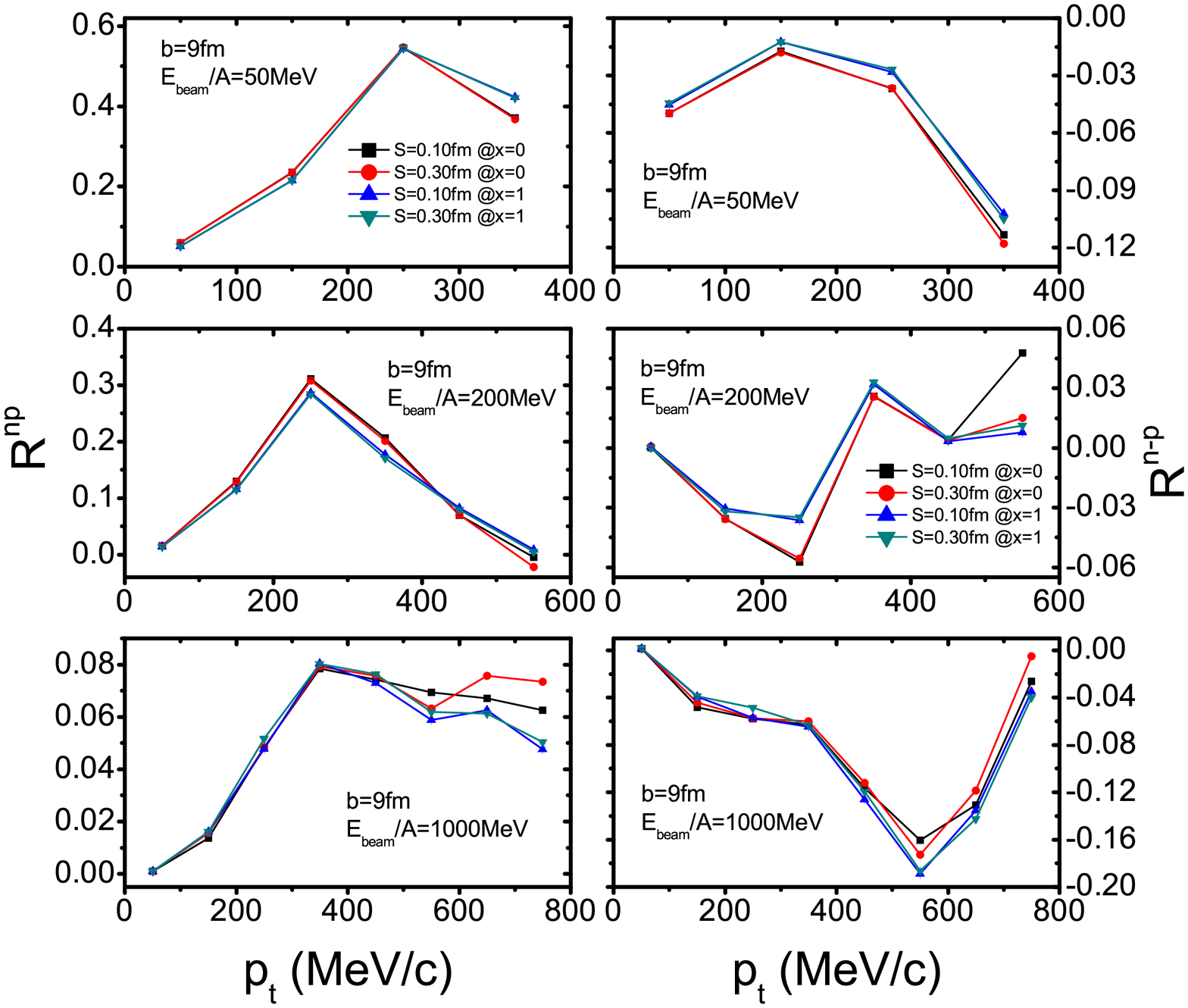}} \caption{(Color
online) Same as Fig. \ref{npsp-pt1} but with impact parameter of 9fm.} \label{npsp-pt2}
\end{figure}

Let's first look at the global dependence of the stopping power on beam energy, impact
parameter and symmetry energy for a given neutron-skin thickness of 0.1fm. Shown in
Fig. \ref{sp-global} is the beam energy and impact parameter dependence of stopping power
of all nucleons in $^{208}$Pb+$^{208}$Pb collisions with the stiff and soft symmetry energy,
respectively. It is seen the stopping power of all nucleons is decreasing with the beam energy
and impact parameter increasing regardless of the stiff or soft symmetry energy. These are
consistent with previous results in Ref. \cite{Luo07,Leh10}. However, the stopping power
of all nucleons is not sensitive to symmetry energy at all. Is the stopping power of neutrons
or protons sensitive to symmetry energy? Can the combined stopping power
eliminate the influence of neutron-skin size difference of initial colliding nuclei but keep the
effect of symmetry energy? To answer these questions, let's first define the neutron-proton
differential stopping power $R^{np}$ and neutron-proton stopping power difference $R^{n-p}$
similar to those of combined collective flow as follows,
\begin{eqnarray}\label{Rnp}
R^{np}(u)=\frac{N_{n}(u)}{N(u)}<R_{n}(u)>-\frac{N_{p}(u)}{N(u)}<R_{p}(u)>,
\end{eqnarray}
\begin{eqnarray}\label{Rnp}
R^{n-p}(u)=<R_{n}(u)>-<R_{p}(u)>,
\end{eqnarray}
where $N(u)$, $N_{n}(u)$ and $N_{p}(u)$ are the numbers of nucleons,
neutrons and protons at parameter $u$ which denotes the rapidity $y$
or transverse momentum $p_{t}$, respectively. Shown in Figs. \ref{nsp-ry}, \ref{psp-ry} and
\ref{npsp-ry} are the rapidity dependence of stopping power of neutrons, protons and
neutron-proton differential stopping power in $^{208}$Pb+$^{208}$Pb collisions with impact
parameters of 5 and 9fm and at beam energies from 50 to 1000MeV/nucleon, respectively. It can
be found that although the stopping power of neutrons and/or protons does not show obvious
sensitivities to symmetry energy, the neutron-proton differential stopping power shows obvious
sensitivities to symmetry energy especially for midrapidity nucleons and at the lower beam energy.
In addition, the neutron-proton stopping power is hardly affected by the neutron-skin size
difference of initial colliding nuclei. This is very similar to that of combined transverse
and elliptic flows due to their correlation as pointed out in Ref. \cite{And06}, i.e., flow
is generated by pressure gradients established in compressed matter, while the achieved
density is connected to the degree of stopping.

Finally, noticing that the midrapidity neutron-proton differential stopping power is obvious
sensitive to symmetry energy but hardly affected by the neutron-skin size difference of
initial colliding nuclei, the transverse momentum dependence of midrapidity neutron-proton
combined stopping power including the neutron-proton differential stopping power
and neutron-proton stopping power difference is shown in Figs. \ref{npsp-pt1} and \ref{npsp-pt2},
respectively. It is shown that the transverse momentum dependence of midrapidity neutron-proton
combined stopping power shows more obvious sensitivities to symmetry energy but hardly affected
by the neutron-skin size difference especially at the lower beam energy.

\section{Summary}
\label{summary}
In this work by studying the influence of the neutron-skin size of initial colliding nuclei
on the collective flow and nuclear stopping power, we have showed how to eliminate the
influence of the neutron-skin size difference of initial colliding nuclei in semicentral
and peripheral Pb+Pb collisions at beam energies from 50 to 1000 MeV/nucleon, respectively.
Noticing that the effects of neutron-skin size on collective flow of neutron and proton are
approximately identical except for the additional Coulomb repulsion between protons, it is
thus that combination of neutron and proton collective flows, i.e., neutron-proton differential
transverse and elliptic flows and neutron-proton transverse and elliptic flow differences, can
effectively eliminate the effects of neutron-skin size difference especially at the lower
beam energy and thus can be as useful sensitive observables of nuclear matter symmetry energy
in heavy-ion collisions. In addition, the combined stopping power including neutron-proton
differential stopping power and neutron-proton stopping power difference also shows some
sensitivities to symmetry energy but hardly affected by neutron-skin size difference of initial
colliding nuclei especially at the lower beam energy.\\

\noindent{\textbf{Acknowledgements}} \\
The author is grateful to Profs. B.A. Li, J.Xu and L.W. Chen for their helpful discussions.
The author also thanks the help provided by the supporting staff of the High-Performance Computational
Science Research Cluster at Texas A$\&$M University-Commerce where partial calculations were done.
This work was supported by the National Natural Science Foundation of China under grant No.11405128.

\end{document}